\documentclass[aps,prb,superscriptaddress,twocolumn,showpacs,showkeys]{revtex4}
\usepackage{graphicx}
\usepackage{amsmath}

\begin{document}

\title{Spin glass properties of an Ising antiferromagnet on the Archimedean $(3,12^2)$ lattice}

\author{M. J. Krawczyk}
\email{krawczyk@novell.ftj.agh.edu.pl}
\author{K. Malarz}
\homepage{http://home.agh.edu.pl/malarz/}
\author{B. Kawecka-Magiera}
\email{basia@novell.ftj.agh.edu.pl}
\author{A. Z. Maksymowicz}
\email{amax@agh.edu.pl}
\author{K. Ku{\l}akowski}
\email{kulakowski@novell.ftj.agh.edu.pl}
\affiliation{
Faculty of Physics and Applied Computer Science,
AGH University of Science and Technology\\
al. Mickiewicza 30, PL-30059 Krak\'ow, Poland
}

\date{\today}

\begin{abstract}
We investigate magnetic properties of a two-dimensional periodic structure with Ising spins
and antiferromagnetic nearest neighbor interaction. The structure is topologically equivalent 
to the Archimedean $(3,12^2)$ lattice. The ground state energy is degenerate. 
In some ground states, the spin structure is translationally invariant, with the same configuration in each unit cell.
Numerical results are reported on specific heat and static magnetic susceptibility against temperature. 
Both quantities show maxima at temperature $T>0$.
They reveal some sensitivity on the initial state in temperatures where the 
Edwards--Anderson order parameter is positive.
For zero temperature and low frequency of the applied field, the magnetic losses are negligible.
However, the magnetization curve displays some erratic behavior due to the metastable states.
\end{abstract}



\maketitle

\section{Introduction}
\begin{figure}
\begin{center}
\includegraphics[angle=180,width=.45\textwidth]{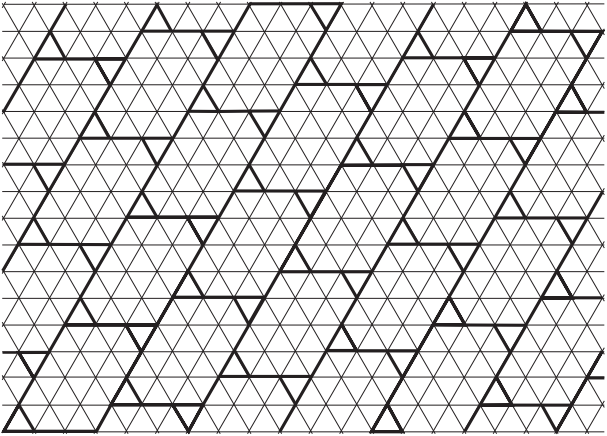}
\caption{The stretched Archimedean $(3,12^2)$ lattice, with the triangular lattice 
as a background.}
\label{lattice}
\end{center}
\end{figure}

The problem of spin glass, with quenched disorder and frustration as main ingredients, 
is known to be a challenge in statistical mechanics.\cite{ns03}
In order to reduce its complexity, it makes sense to discuss these ingredients separately. The aim of 
this work is to report numerical results on magnetic properties of a structure 
where frustration is present for purely antiferromagnetic interaction, with
only one value of the exchange interaction.
The interaction is limited to nearest 
neighbors. The structure studied is presented in Fig. \ref{lattice}. 

\begin{figure}
\begin{center}
\includegraphics[width=.1\textwidth]{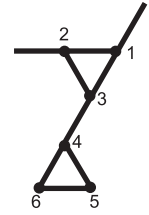}
\caption{A unit cell.}
\label{cell}
\end{center}
\end{figure}

Our motivation to discuss this structure is as follows. First, the frustration is to be present,
and the simplest way to achieve it is to put triangles into the lattice. Second, the number of 
nearest neighbors is to be odd, what ensures that for any magnetic state, the energy
barrier to flip a spin is finite. Then we may expect that the magnetic transition 
temperature is greater than zero.  Third, we are interested in the ground state degeneracy.
We look for a lattice where, if external field is zero, a simultaneous flip of some
spins does not change the total energy. In fact, if the magnetization of the flipped group 
of spins is zero, the flipping does not change the energy even in the presence of the applied 
field. Fourth, all distances between the nearest neighbors should be equal. Additionally,
two-dimensional structures are preferred for their simplicity. The structure 
presented in Fig. \ref{lattice} fulfills all these conditions. The last condition is guaranteed by 
spanning our structure on the triangular lattice. Actually, this structure
is topologically equivalent to the Archimedean lattice $(3,12^2)$. Patterns 
of all eleven Archimedean lattices can be found in Ref. \onlinecite{sz}.
Topological equivalence means that one lattice can be stretched into the other.
Magnetism of Archimedean lattices has been investigated for some years.\cite{tom,rich}
However, these works deal with the isotropic Heisenberg interaction.
We are not aware about temperature-dependent simulations on the $(3,12^2)$ lattice with Ising antiferromagnetism. 

\section{The ground state}

The energy of the classical Ising state\cite{ising} in the presence of external magnetic field $H$ is
\begin{equation}
E=-\dfrac{1}{2}J\sum_{(i,j)}S_iS_j-H\sum_iS_i
\end{equation}
where $S_i=\pm 1$, $J<0$ is the antiferromagnetic exchange constant and the first summation goes over all nearest neighbor pairs $(i,j)$. The degeneracy of the ground state energy (GSE) of the structure 
presented in Fig. \ref{lattice}, termed as stretched Archimedean
 for brevity from now on, can be demonstrated easily when we look at the unit cell in Fig. \ref{cell}. 
There are nine bonds in the unit cell. The ground state energy cannot be smaller than it is allowed 
by the frustration, which is unavoidable within the two triangles. The contribution from the triangles 
to GSE is then $2J$ per unit cell. Provided that the spins connected with other bonds are 
oriented in antiparallel, we get $5J$ per unit cell for field equal to zero. This is well possible,
and some the ground states are periodic. One of such states is
$(s_1,s_2,s_3,s_4,s_5,s_6)=(-,+,+,-,-,+)$. Another periodic 
ground state can be obtained by a simultaneous flip of spins $s_3,s_4$ in each unit cell. 
More general, we have six periodic ground states $(s_1,s_2,s_3,-s_3,-s_2,-s_1)$; additional
condition is that $s_1+s_2+s_3=\pm 1$. These states are collected in Tab. \ref{tab1}. 

\bigskip
\begin{table}
\caption{List of homogeneous ground states, with spins labeled as in Fig. \ref{cell}. 
In last column, paths are indicated which lead to other states by flipping pairs of spins in all unit cells.}
\label{tab1}
\begin{ruledtabular}
\begin{tabular}{ccc}
Ground &                            &         Pairs \\
state  &  $s_1,s_2,s_3,s_4,s_5,s_6$ & which can be flipped \\
\hline
A & $-,+,+,-,-,+$ & ($s_3, s_4$) to C or ($s_2, s_5$) to B \\
B & $-,-,+,-,+,+$ & ($s_2, s_5$) to A or ($s_1, s_6$) to F \\
C & $-,+,-,+,-,+$ & ($s_1, s_6$) to E or ($s_3, s_4$) to A \\
D & $+,-,-,+,+,-$ & ($s_3, s_4$) to F or ($s_2, s_5$) to E \\
E & $+,+,-,+,-,-$ & ($s_2, s_5$) to D or ($s_1, s_6$) to C \\
F & $+,-,+,-,+,-$ & ($s_1, s_6$) to B or ($s_3, s_4$) to D \\
\end{tabular}
\end{ruledtabular}
\end{table}

Obviously, many other non-periodic ground states of the whole lattice can be obtained, 
for example from $(-,+,+,-,-,+)$ if the flip
of two spins is performed only in a selected number of unit cells. This means that the 
ground state degeneracy increases with the lattice size at least as $6\cdot 2^{N/6}$, 
where $N$ is the number of sites.

\section{Numerical results}

\begin{figure}
\begin{center}
\includegraphics[angle=-90,width=.45\textwidth]{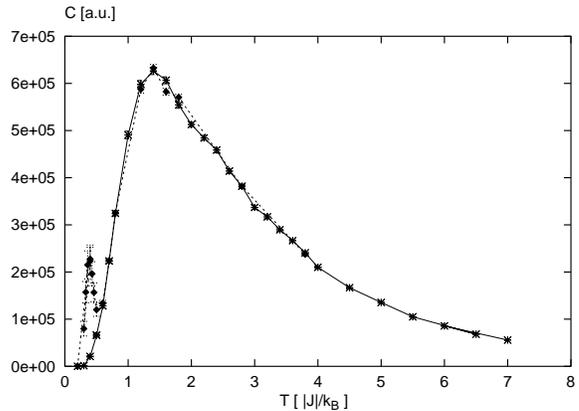}
\caption{Temperature dependence of the specific heat $C$. We show two curves: for saturated (solid
line) and random (dotted line) initial states. At low temperatures, i.e. below $T=0.7~[|J|/k_B]$, 
the results depend on the initial state. However, the main maximum of $C(T)$ is the same 
for both initial states.}
\label{cv}
\end{center}
\end{figure}

We apply the heat bath Monte Carlo approach\cite{herr} to find the magnetic contribution to the 
specific heat at zero field, i.e. 

\begin{equation}
C=\beta (\langle E^2\rangle-\langle E\rangle^2),
\end{equation}
where $\langle\cdots\rangle$ is an average over thermodynamic ensemble, and $\beta =1/(k_BT)$. In the
numerical calculations, the thermal average $\langle\cdots\rangle$ is substituted by the time average. 
A lattice of $6\cdot 10^4$ spins
is used, with periodic boundary conditions. The initial state is of full saturation, i.e.  all spins equal to $+1$.
Alternatively spins are randomly +1 or $-1$, with equal weights.
After about $100$ time steps, the total magnetization is close to zero,
and the system is reasonably close to thermal equilibrium, at least for high temperatures.
We define a time step as one update of the whole network.
Then,
the time average of $E$ and $E^2$ is found. The results are averaged over $N_{\text{run}}=100$ trials. 
For $T$ less than $0.7~[|J|/k_B]$ the statistics is better: $N_{\text{run}}=10^3$.
The obtained plot is shown in Fig. \ref{cv}.

\begin{figure}
\begin{center}
\includegraphics[angle=-90,width=.45\textwidth]{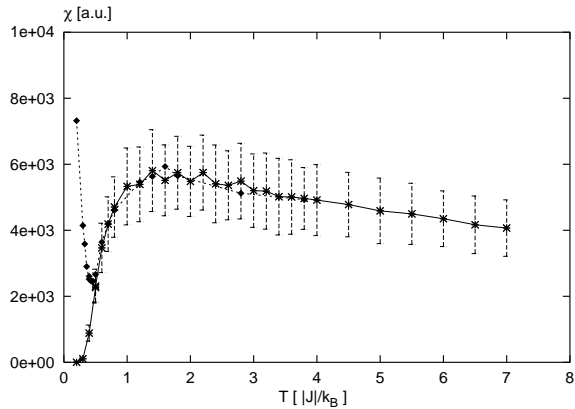}
\caption{Thermal dependence of the static spin susceptibility $\chi$ for zero field. The results
for the saturated initial state (solid line) show that below the maximum, $\chi$ tends mildly to
zero, with numerical uncertainties decreasing at low $T$. The results for the random initial state 
(dotted curve) show irregularities below $T=0.7~[|J|/k_B]$. In this range of temperature, we observe a 
rapid increase of $\chi$ below the main maximum, and numerical uncertainties increase when $T$ is 
lowered. For $T=0.2~[|J|/k_B]$, the uncertainties are the largest and reach about 100 percent of the obtained 
value. }
\label{chi}
\end{center}
\end{figure}

The same algorithm is used to calculate the static susceptibility for zero field, i.e.

\begin{equation}
\chi=\beta^2 (\langle M^2\rangle-\langle M\rangle^2),
\end{equation}
where $M=\sum_iS_i$. Here again, two sets of data $\chi (T)$ are obtained for two different kinds
of the initial conditions, one saturated and one random. The results are shown in Fig. \ref{chi}.
As we see, two curves $\chi (T)$ become different below certain temperature. The curve 
for the random initial state is higher. This is so since, for low temperature, the algorithm
of heat bath leads the saturated system to one of its ground states, whereas the randomness of 
initial state is to some extent preserved.

\begin{figure}
\begin{center}
\includegraphics[angle=-90,width=.45\textwidth]{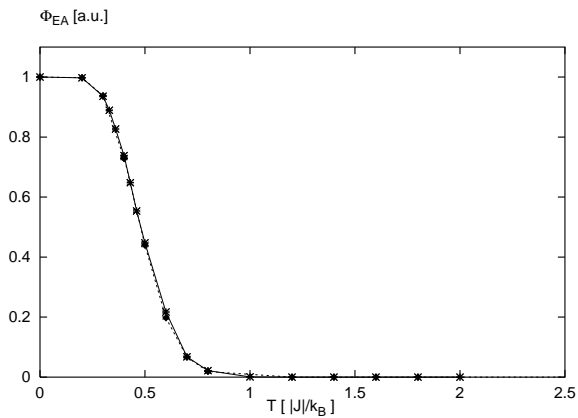}
\caption{Thermal dependence of the Edwards--Anderson order parameter. The results are 
practically the same for initial random state and initial saturated state.}
\label{qEA}
\end{center}
\end{figure}

To get more insight into this memory effect, we also calculate the thermal dependence of the 
Edward--Anderson order parameter $\Phi_{EA}$.\cite{ea}
In its definition below, the appropriate time average is written explicitly:
\begin{equation}
\Phi_{EA}=\sum_i\left(\dfrac{1}{\tau}\sum_{t=1}^\tau S_i(t)\right)^2.
\end{equation}
In Fig. \ref{qEA} we show $\Phi_{EA}(T)$ obtained from a random and saturated states by the heat bath algorithm at zero
temperature. As we see, $\Phi_{EA}$ starts to differ from zero at temperature close to $T\approx 0.7~[|J|/k_B]$, 
where the plots on $C(T)$ and $\chi (T)$ obtained for different initial conditions separate.

\begin{figure}
\begin{center}
(a) \includegraphics[width=.45\textwidth]{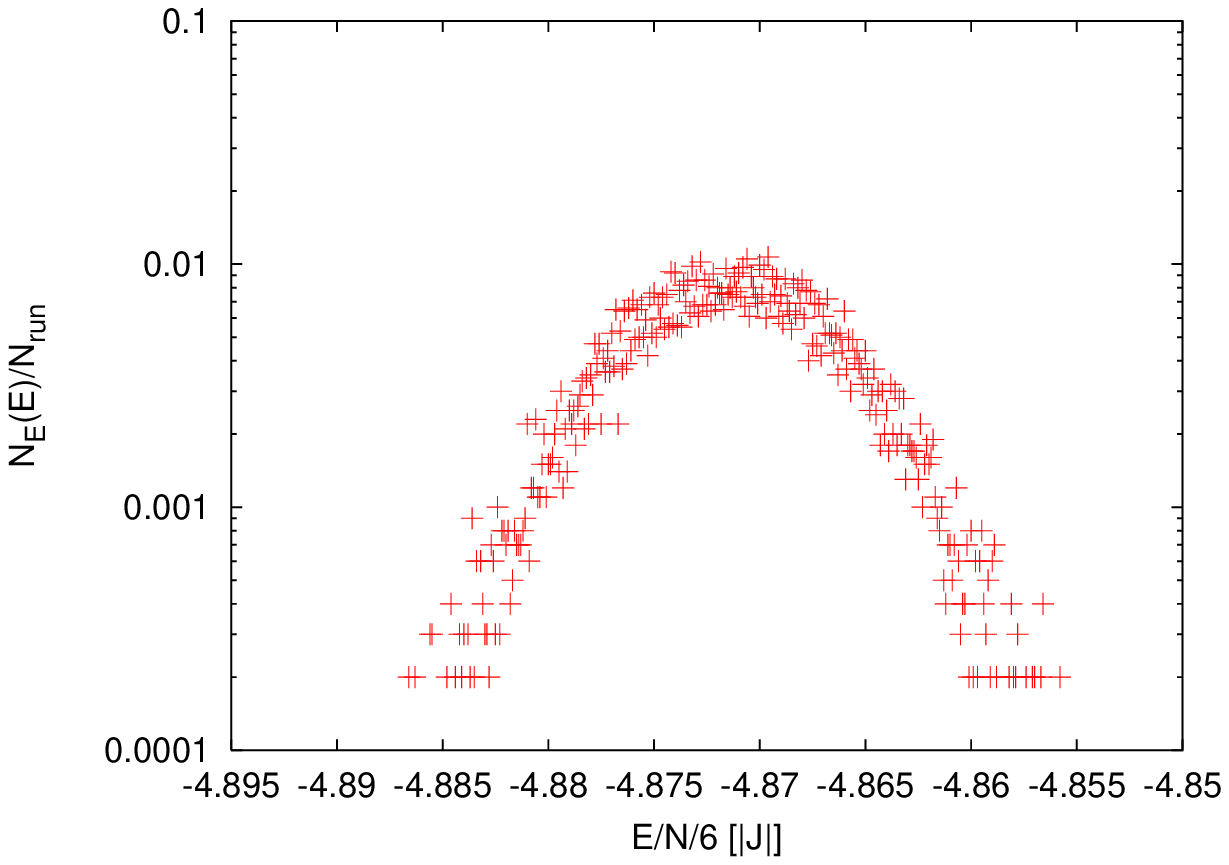}\\
(b) \includegraphics[width=.45\textwidth]{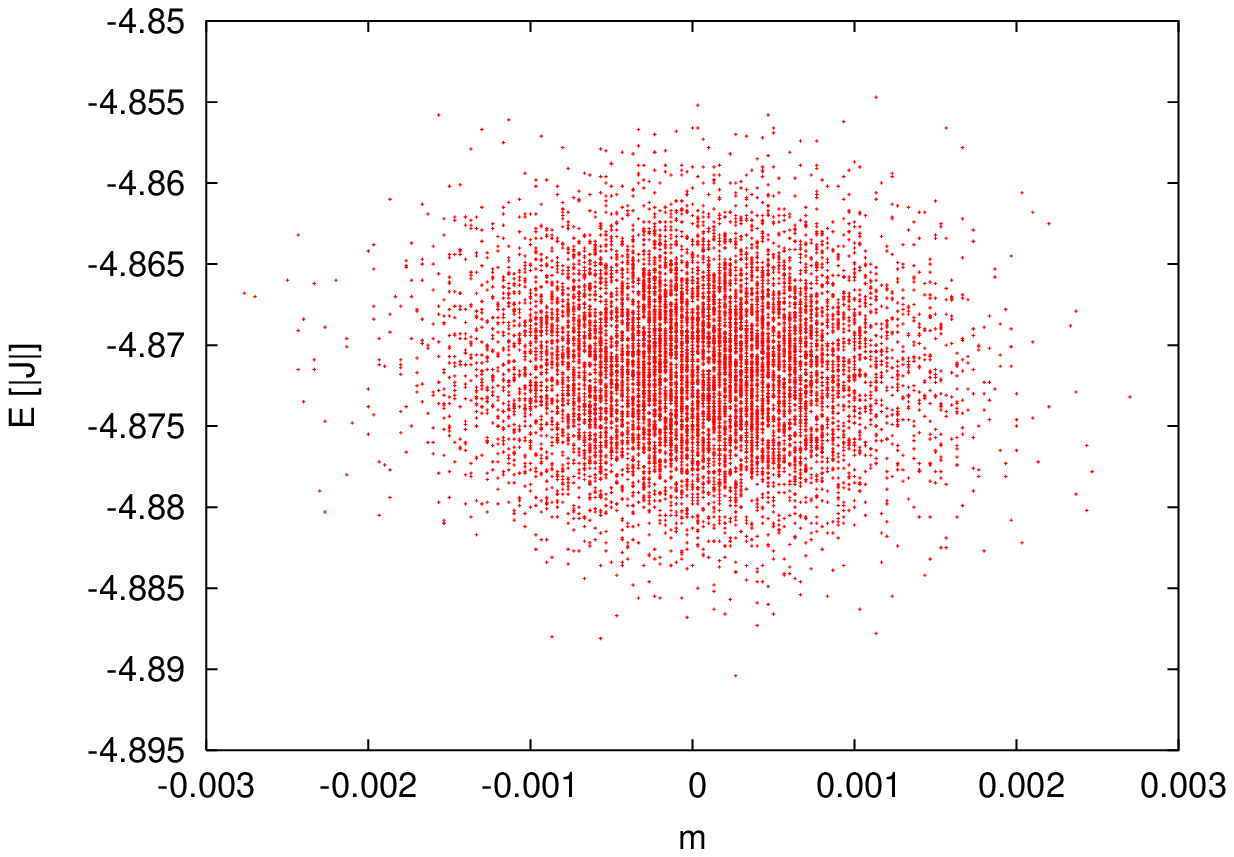}
\caption{(a) Energy distribution of metastable states obtained for $T=0$ from random initial states.
The mean energy is about 2.5 \% higher that GSE=$5J$ per unit cell.
(b) Energy against the reduced magnetization $m=M/N$ of the same metastable states. No visible correlation is found.
$m(t=0)=0$, $H=0$, $T=0$, $N=6\cdot 10^4$, $N_{\text{run}}=10^4$.
}
\label{hist_e}
\end{center}
\end{figure}

For the case of $T=0$, we calculate also the spectrum of the metastable states. These 
states are obtained from a random initial state by means of the heat bath algorithm in zero temperature limit.
The result is shown in Fig. \ref{hist_e}(a). It is seen that there is some randomness preserved in the metastable states.
The mean energy of a metastable state is less than three percent above the ground state energy. In Fig. \ref{hist_e}(b), an attempt is presented to find a correlation
between energy and magnetization of the same metastable states. As we see, there is no correlation 
at all. We note only that metastable states with non-zero magnetic moments do exist and can
produce some contribution to the hysteresis loop. However, this contribution is very small, as seen 
in the next figure.

\begin{figure}
\begin{center}
\includegraphics[width=.45\textwidth]{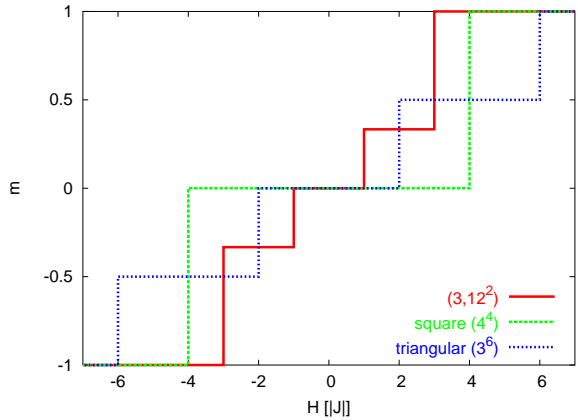}
\caption{The magnetization curves for $(3,12^2)$, the square lattice and the triangular lattice (solid, dashed and dotted lines, respectively) for $T=0$.}
\label{loops}
\end{center}
\end{figure}

\begin{figure*}
\begin{center}
(a) \includegraphics[width=.45\textwidth]{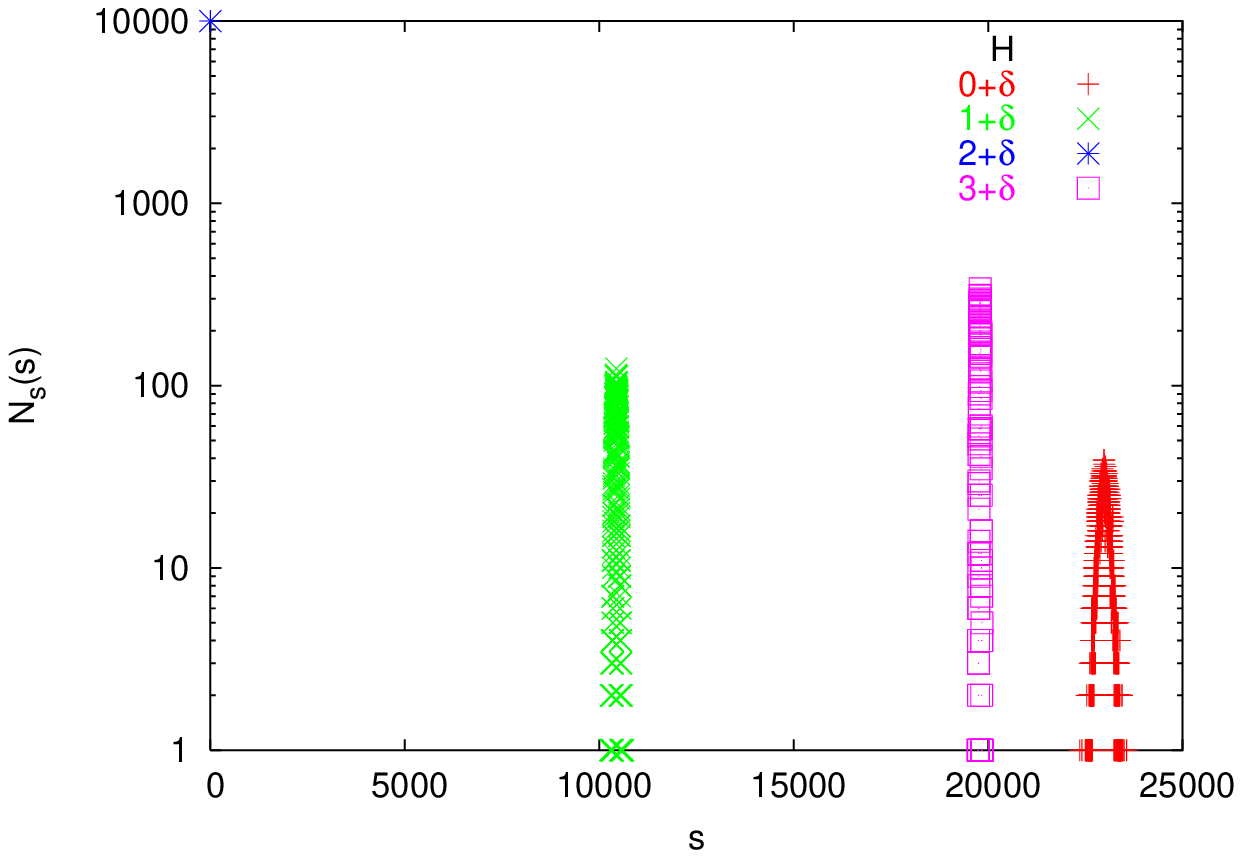}
(b) \includegraphics[width=.45\textwidth]{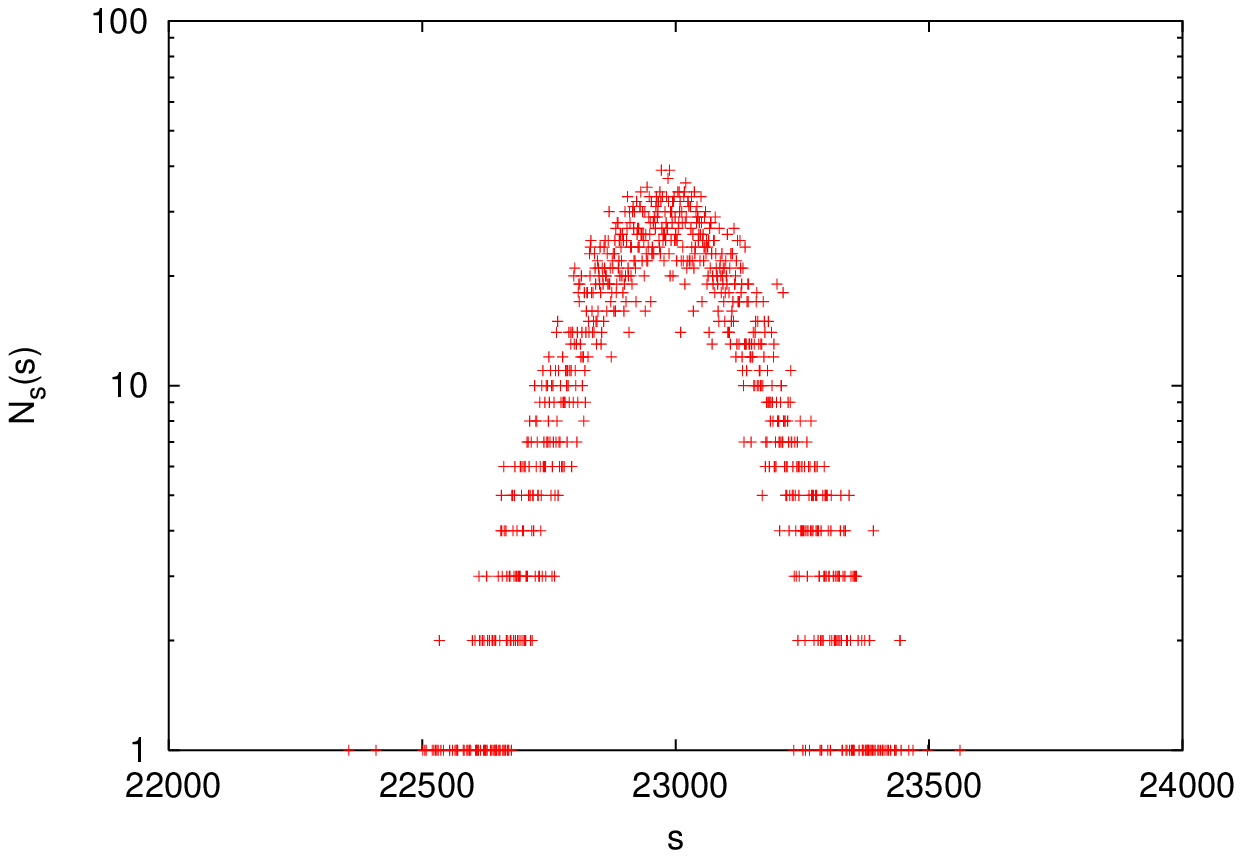}\\
(c) \includegraphics[width=.45\textwidth]{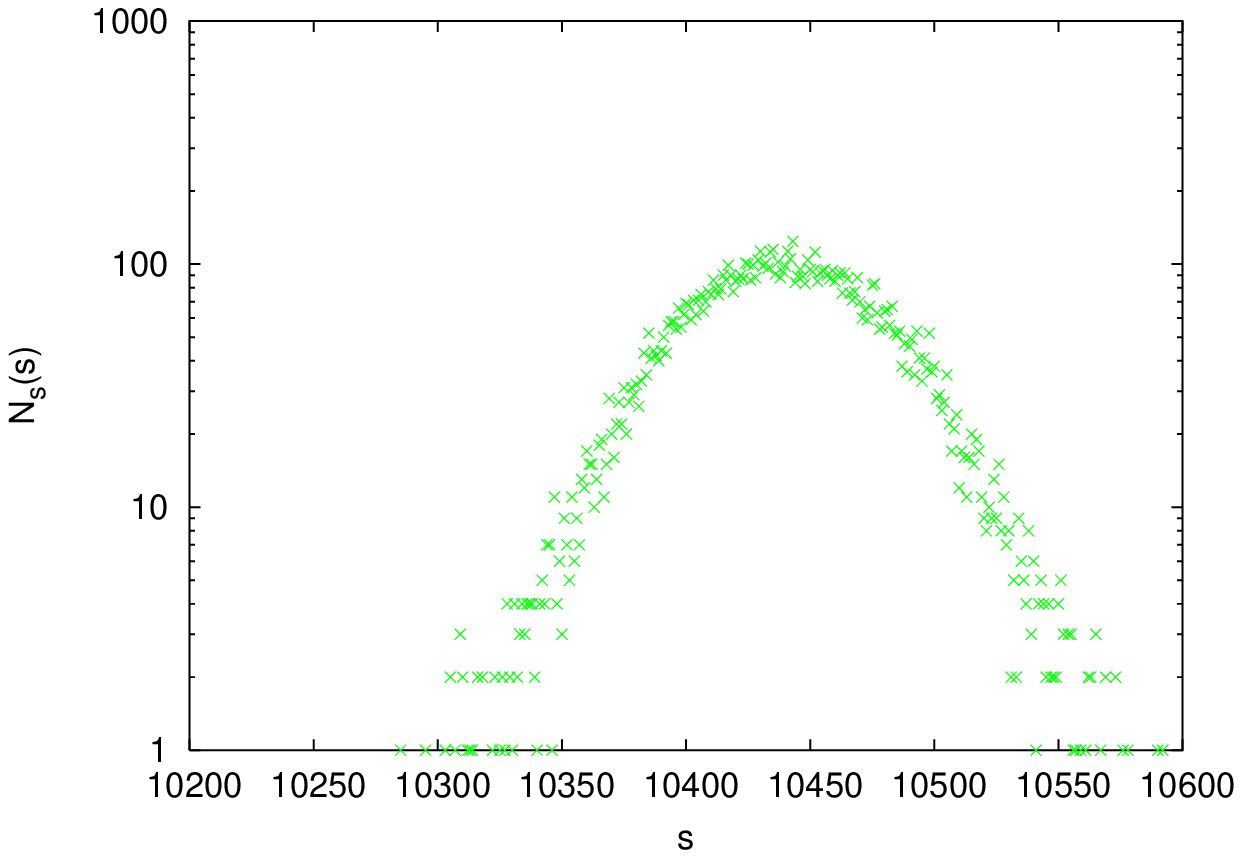}
(d) \includegraphics[width=.45\textwidth]{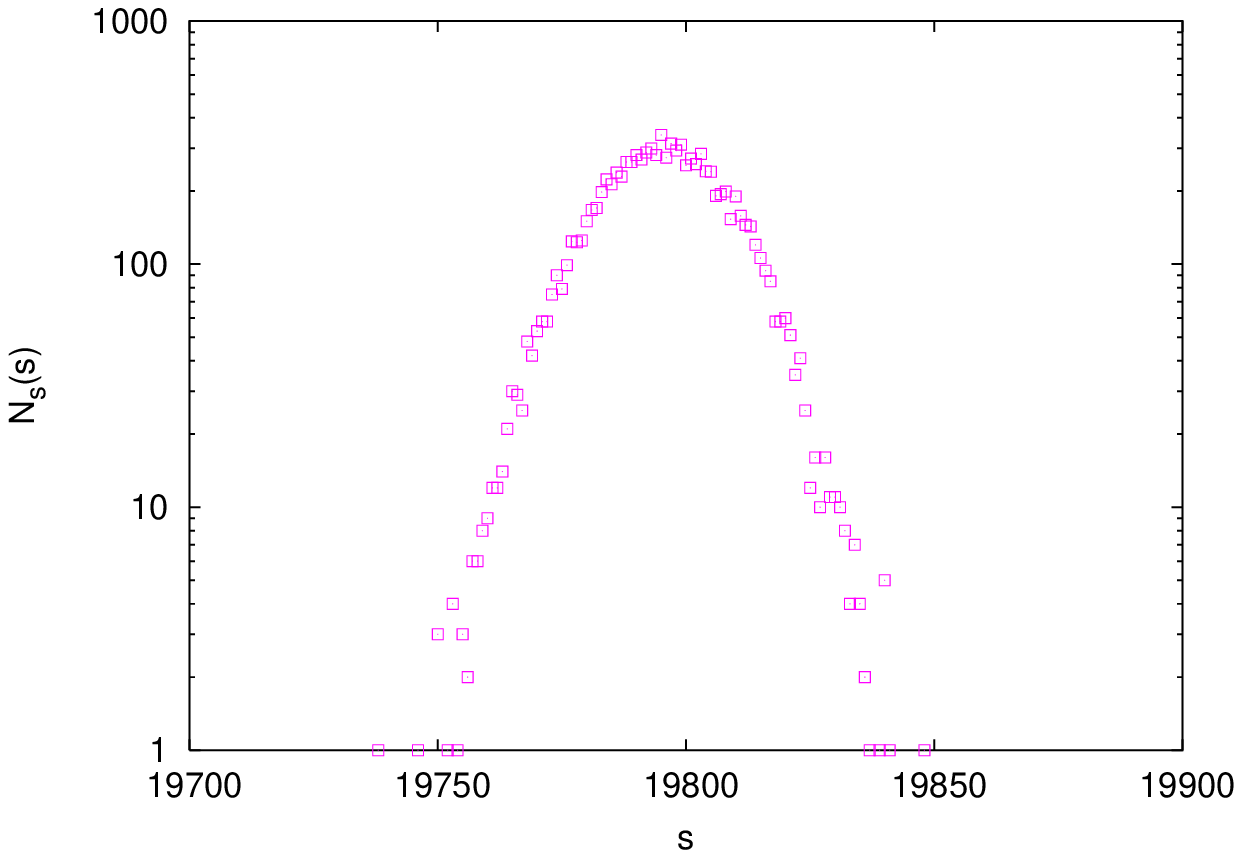}
\caption{The spectrum of avalanches. (a) Three peaks are obtained for avalanches at three different fields: (b) $H=0$, (c) $H=1~[|J|]$, (d) $H=3~[|J|]$.
This result reflects the distribution of the interaction field.
$T=0$, $N=6\cdot 10^4$, $N_{\text{run}}=10^4$.}
\label{peaks}
\end{center}
\end{figure*}

The same kind of randomness is present in the results on the magnetization curve. The curve is shown in Fig. \ref{loops}.
For a comparison, we present also the data obtained for the square lattice 
and the triangular lattice. The distribution of sizes of spin flips avalanches is shown in Fig. \ref{peaks}. 
By an avalanche size we mean the number of flipped spins at a given field. In Fig. \ref{peaks}, we see three maxima, obtained at fields $H=0$, $H=1~[|J|]$ and $H=3~[|J|]$.
The avalanches at $H=0$ are just numbers of spins which flip when the system passes from random initial state to a metastable state, with final energy distribution shown in Fig. \ref{hist_e}(a).
These avalanches are the largest, as they contain about 23000 flippings.
The avalanches occurring at field $H=1~[|J|]$ and $H=3~[|J|]$ contain 10450 and 19800 flippings in the average (see Fig. \ref{peaks}(a)).

Above $H=3~[|J|]$, all spins are saturated.
Provided that, in the average, the magnetization of a metastable state is approximately zero, a half of spins ($3\cdot 10^4$) are to be flipped to saturate the sample.
In Fig. \ref{peaks}(b-d) we show the same peaks in smaller scales, what enables to observe their detailed character.
The positions of the peaks of the spectrum indicate, that the summarized size of avalanches at $H=1~[|J|]$ and $H=3~[|J|]$ is 30250 in the average.
The small amount of difference, here 250 spins in the average,
mean that some spins flip back and forth. 

\section{Discussion}

The magnetic properties of the investigated lattice shows similarities to the spin glass
state. These are: maxima of the spin susceptibility $\chi$ and the specific heat $C$ as dependent 
on temperature, and the memory effect, i.e. the dependence of these observables on an 
initial state below $T=0.7~[|J|/k_B]$.  In this range of $T$, 
the Edwards--Anderson order parameter $\Phi_{EA}$ is different from zero. We deduce that there,
the ensemble average cannot be substituted by the time average, and our numerical results on $C$
and $\chi$ are not reliable. However, the main peaks of $C$ and $\chi$ between $T=1.0~[|J|/k_B]$ and $T=2.0~[|J|/k_B]$
do not depend on the initial state.

On the other hand, the ground state energy is multiply degenerated. As a consequence, the overlap\cite{ns03} of two ground states $\alpha$ and $\gamma$, defined as
\begin{equation}
q_{\alpha \gamma}=\frac{1}{N}\sum_i^N S_i(\alpha)S_i(\gamma)
\end{equation}
is different from unity.
For example, the overlap between states A and B from Tab. \ref{tab1} is $q_{AB}=1/3$. When the non-periodic ground states are taken into account, the overlap distribution $P(q)$ is practically a continuous function 
of $q$. To explain it, let us consider only two periodic ground states, A and B. If only 
those two are possible, an overlap between two non periodic ground states $\alpha$ and $\gamma$ 
within a unit cell is either $1$ or $1/3$, for the cells in the same or different states,
respectively. For any state $\alpha$, we can select $n_s$ out of $N/6$ unit cells and form 
state $\gamma$ putting them in the same states, as in $\alpha$. In this case we have the 
overlap probability 
\begin{equation}
P\Big(q=\frac{n_s+(N/6-n_s)/3}{N/6}\Big)=\frac{(N/6)!}{n_s!(N/6-n_s)!}.
\end{equation}

In fact, there are six periodic ground states, and not only two. Information, cells in which 
ground states can be neighbors without raising energy, is given in the last column of Tab. \ref{tab1}.
If the condition of periodicity is removed, the number of ground states of a unit cell increases.
In any case, a typical ground state of the whole lattice is expected to be not periodic,
but disordered. In this way, the most simple definition of spin glass\cite{rev} is true for our lattice.  
In the case of $T=0$, the system dynamics leads to (meta)stable states, which depend on initial states.
The spectrum of GSE and the magnetization curve reveal a random character, which is a consequence of the random initial state.

Numerical results reported above suggest, that the temperature of the transition between the paramagnetic state and the low-temperature state, which seems to be the spin-glass state, is positive.
It is an open question, how much disorder is needed to reproduce the vanishing of the transition temperature, which is the standard result for two-dimensional
Ising spin glasses.\cite{rev}
With small amount of disorder, we expect that the interaction field at some sites is zero and these spins flip freely.
Once a cluster of these spins spans throughout the lattice, the transition at finite temperatures is likely to disappear.

Concluding, the $(3,12^2)$ Archimedean lattice can be useful as a reference example when the question, which features of a realistic spin glass are a consequence of frustration only, is under debate.
This structure can be also a good starting point for numerical calculations.
Last but not least, it is much easier to simulate an Ising spin glass where the magnetic interaction is homogeneous.

\begin{acknowledgments}
Part of calculations was carried out in ACK\---CY\-FRO\-NET\---AGH.
The machine time on HP Integrity Superdome is financed by the Polish Ministry of Science and Information Technology under grant No. MNiI/\-HP\_I\_SD/\-AGH/\-002/\-2004.
\end{acknowledgments}



\begin{thebibliography}{88}
\bibitem{ns03}
C. M. Newman and D. L. Stein,
J. Phys.: Condens. Matt. {\bf 15} (2003) R1319.
\bibitem{sz}
P. N. Suding and R. M. Ziff,
Phys. Rev. E {\bf 60} (1999) 275.
\bibitem{tom}
P. Tomczak and J. Richter,
Phys. Rev. B {\bf 59} (1999) 107.
\bibitem{rich}
J. Richter, J. Schulenburg and A. Honecker,
{\it Quantum Magnetism in Two Dimensions: From Semi-classical N\'eel Order to Magnetic Disorder},
Lect. Notes Phys. {\bf 645} (2004) 85.
\bibitem{ising}
W. Lenz,
Z. Phys. {\bf 21} (1920) 613;

E. Ising,
Z. Phys. {\bf 31} (1925) 253.
\bibitem{herr}
D. W. Heermann,
{\it Computer Simulation Methods in Theoretical Physics},
(Springer-Verlag, Berlin, 1990).
\bibitem{ea}
S. F. Edwards and P. W. Anderson,
J. Phys. F: Metal Phys. {\bf 5} (1975) 965.
\bibitem{rev}
K. H. Fischer and J. A. Hertz,
{\it Spin Glasses},
(Cambridge UP, Cambridge, 1991).
\end{thebibliography}
\end{document}